\documentclass[sigconf]{acmart}
\AtBeginDocument{%
  \providecommand\BibTeX{{%
    \normalfont B\kern-0.5em{\scshape i\kern-0.25em b}\kern-0.8em\TeX}}}

\setcopyright{acmcopyright}
\copyrightyear{2022}
\acmYear{2022}
\usepackage{enumitem}
\usepackage{algorithmic}
\usepackage{graphicx}
\usepackage{subcaption}
\usepackage{textcomp}
\usepackage{xcolor}
\usepackage{bussproofs}
\definecolor{dkgreen}{rgb}{0,0.5,0}
\definecolor{dkred}{rgb}{0.5,0,0}
\definecolor{dkgray}{rgb}{0.3,0.3,0.3}
\usepackage{balance}
\usepackage{listings}
\usepackage{xspace}
\usepackage{tablefootnote}
\usepackage{booktabs} 

\newcommand{\tool}{\texttt{COOLIO}}

\begin{document}

\author{Linhan Li}
\affiliation{
\institution{George Mason University}
  \country{USA}
}

\author{ThanhVu Nguyen}                                           
\affiliation{
\institution{George Mason University}
  \country{USA}
}

\title{\tool{}: A Language Support Extension for the \\
Classroom Object Oriented Language}

\begin{abstract}
COOL is an Object-Oriented programming language used to teach compiler design in many undergraduate and graduate courses.
Because most students are unfamiliar with the language and code editors and IDEs often lack the support for COOL, writing code and test programs in COOL are a burden to students, causing them to not fully understand many important and advanced features of the language and compiler.
In this tool paper, we describe \tool{}, an extension to support COOL in the popular VSCode IDE. \tool{} provides (i) syntax highlighting supports for the COOL language through lexing and parsing, (ii) semantics-aware autocompletion features that help students write less code and reduce the burden of having to remember unfamiliar COOL grammar and syntax, and (iii) relevant feedback from the underlying COOL interpreter/compiler (e.g., error messages, typing information) to the students through VSCode editor to aid debugging.  We believe that \tool{} will help students enjoy writing COOL programs and consequently learn and appreciate more advanced compiler concepts. 
\end{abstract}
\maketitle

\section{Introduction}\label{sec:intro}

COOL (Classroom Object Oriented Language) is a language often used in undergraduate or graduate compiler classes. These include traditional courses at Stanford, UC-Berkeley, University of Nebraska–Lincoln, University of Virginia, University of Michigan, Vanderbilt,a nd online ones at Coursera, EdX, and Stanford Online. 
The language is sufficiently small for students to implement a complete interpreter or compiler for it within a semester, but is still powerful enough to support major concepts in a modern programming language including polymorphism, inheritance, dynamic dispatching, type checking, and automatic garbage collection.
As described by Alex Aiken, its author, the COOL language is "object-oriented, statically typed, and has automatic memory management"~\cite{Aiken1996CoolAP}.

While being the chosen language for many Compilers and Programming Languages courses, the syntax and semantics of COOL are unfamiliar to students. 
Thus, students often just write simple COOL programs to test their Compiler implementation. Such simple tests cause students to miss many important features of language and therefore miss crucial designs of the compiler to handle them (e.g., dynamic dispatching, inheritance, complex memory usage).  The reason is not that students are not willing to create more complicated and thorough tests, but because they are not very excited to write long and complex COOL programs without help from their favorite code editors.

\paragraph{Related Work} Currently, Extensions for editors and IDEs to support COOL, e.g., 
\texttt{sublime-cool-highlighter}\footnote{\url{https://github.com/princjef/sublime-cool-highlighter}} for SublimeText, \texttt{atom-language-cool\footnote{\url{https://github.com/princjef/atom-language-cool}}} for Atom, and \texttt{language-cool}\footnote{\url{https://marketplace.visualstudio.com/items?itemName=maoguangming.cool}} for VSCode, provide very basic syntax highlighting for COOL keywords. Moreover, they  miss many basic features of syntax highlighting, e.g., the classes/variables defined by the user are not properly highlighted or indented.
Moreover, while syntax highlighting is useful, it is not sufficient to engage users to use COOL.
These extensions do not include important features such as auto-completion and error-reporting, and thus students still need to memorize all the syntax or repeatedly look at the manual or run the provided reference compiler to check for syntax errors.
In fact, because these extensions are not well-maintained (e.g., \texttt{language-cool} has not been updated since 2016)  and do not properly cover even the basics of Cool, most students instead just write COOL programs in plain text mode.

\paragraph{This Work} To help students learn COOL, we have developed \tool{}, a portable,  feature-rich extension for COOL to be used in major IDEs supporting the increasingly popular Language Server Protocol (LSP)~\cite{LSP}.  
While in this paper we demonstrate \tool{} using the VSCode IDE, which is a favorite among students and popular IDE (consistently ranked as one of the top IDE choices), \tool{} works with any IDE supporting LSP such as Atom, Eclipse, Emacs, SublimeText, Vim.

\tool{} supports syntax highlighting for the entire COOL language by using regular expressions from the COOL lexer and parser to recognize COOL syntax. \tool{} also has quick and useful autocompletion features that one would expect from a modern IDE, e.g., code suggestion and auto structure completion, all of which help the student write more correct code and reduce the burden of memorizing COOL syntax and grammar.  Finally, \tool{} exploits LSP client and server design to send COOL's type checking and evaluation information back to the user through the VSCode editor (e.g., underline incorrect lines of code, giving the reason why type checking fails in real-time), allowing the students to quickly understand and debug COOL programs.

We believe that \tool{} can help students in Compiler or Programming Languages courses that use COOL.
This includes our Compiler class, which typically has on average about 30 undergraduate and graduate students, and Compiler and PL courses taught at other places and might have even more students (e.g., at Stanford, UMich, or online courses such as Coursera). We have started using \tool{} ourselves and have already let colleagues teaching courses involving COOL know about \tool{} so that they can refer their students to try the tool. We believe that the rich features provided by \tool{} will be useful to the students, allowing them to learn more about COOL and design a better compiler to support advanced features of the language.

\tool{} is open-source and available at \textbf{\url{https://github.com/dynaroars/COOL-Language-Support}}.
The users can easily search and install it directly from the VSCode IDE or Marketplace\footnote{\url{https://marketplace.visualstudio.com/items?itemName=Linhan.cool-language-support}}.

\section{COOL}

\begin{figure}
\includegraphics[width=0.75\linewidth]{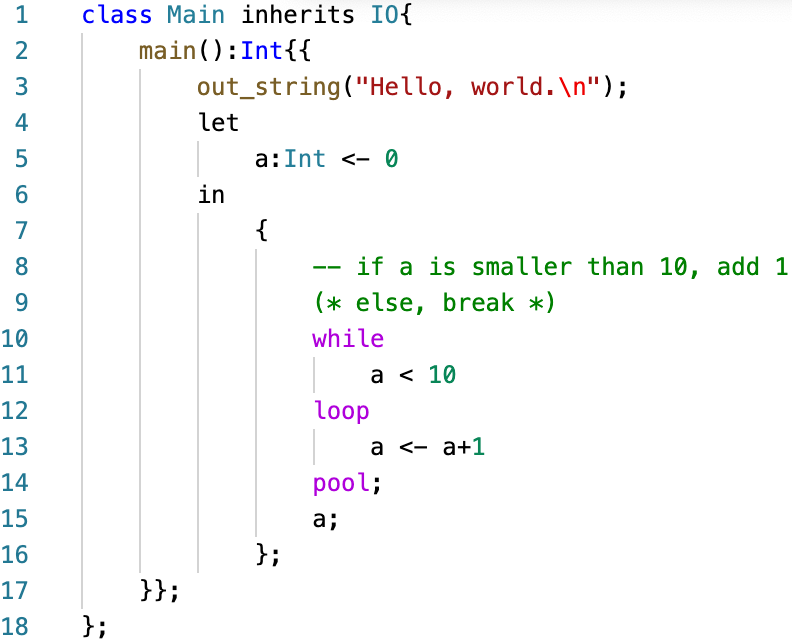}
\caption{Simple Cool program in VSCode with \tool{}}\label{fig:syntax_highlight}
\end{figure}

COOL is an Object-Oriented language and shares some design and syntax with Java. %
Fig.~\ref{fig:syntax_highlight} shows a small COOL program that prints \texttt{HelloWorld} (lines 1-3) and a simple loop.
The COOL manual, called CoolAid~\cite{cool_homepage}, provides the formal syntax and semantics of the language, which we summarize below.

\paragraph{Syntax} The syntax of COOL is designed to be simple. This allows students to write regular expressions (regexes) for lexer and parser to recognize syntactically correct COOL programs. The standard COOL specification does not have advanced features such as list/array structures, threading/multi-processing supports, or exception handling (supports for these are often used as homework or extra credit assignments).

\begin{table}
\caption{Cool Syntax and Grammar}\label{fig:cool_syntax}
\small
\begin{tabular}{ r c l }
\textit{program} & ::= & $[class;]^+$\\
\textit{class}   & ::= & \textbf{class} TYPE [\textbf{inherits} TYPE] $\lbrace$ [\textit{feature;}]$^*\rbrace$ \\
\textit{feature} & ::= & ID([formal[,formal]$^*$]):TYPE $\lbrace$ \textit{expr} $\rbrace$ \\
                 & \texttt{|} & ID:TYPE[<- \textit{expr}] \\
\textit{formal}  & ::= & ID:TYPE \\
\textit{expr}    & ::= & ID <- \textit{expr} \\
            & \texttt{|} & \textit{expr}[@TYPE].ID ([\textit{expr}[\textit{\,expr}$^*$]) \\
            & \texttt{|} &  ID([\textit{expr} [\textit{,expr}]$^*$]) \\
            & \texttt{|} &  \textbf{if} \textit{expr} \textbf{then} \textit{expr} \textbf{else} \textit{expr} \textbf{fi} \\
            & \texttt{|} &  \textbf{while} \textit{expr} \textbf{loop} \textit{expr} \textbf{pool} \\
            & \texttt{|} &  \{[\textit{expr};]$^+$\} \\
            & \texttt{|} &  \textbf{let} ID:TYPE [<- \textit{expr}] [,ID:TYPE [<-\textit{expr}]]$^*$ \textbf{in} \textit{expr}\\
            & \texttt{|} &  \textbf{case} \textit{expr} \textbf{of} [ID:TYPE=> \textit{expr}]$^+$ \textbf{esac} \\
            & \texttt{|} &  \textbf{new} TYPE 
            \texttt{|} \textbf{isvoid} \textit{expr} 
            \texttt{|} \textasciitilde \textit{expr} 
            \texttt{|} \textbf{not} \textit{expr} \\
            & \texttt{|} &  (\textit{expr}) 
                        \texttt{|} \textit{expr [+|-|*|/] expr} 
            \texttt{|} \textit{expr [<|<=] expr} \\
            & \texttt{|} &  ID 
            \texttt{|}  constant(\textit{integer|string|true|false}) \\
\end{tabular}
\end{table}

Tab.~\ref{fig:cool_syntax} displays the syntax and context-free grammar of COOL. 
A COOL program is a set of COOL \texttt{classes}, and each class consists of \texttt{features} which are attributes (variables) and methods. 
Each class defines a type and thus programmer defines new types and associated data and methods by creating new classes (similar to an OO language such as Java).
COOL is an expression language and thus most COOL constructs are expressions.  Expressions take up a large portion of the COOL syntax, but in general, are relatively straightforward and similar to expressions in traditional languages. Note that the \texttt{let} expression that declares a new variable is similar to the one used in a functional language such as OCaml.

\paragraph{Type Checking and Semantics} COOL is a \emph{type-safe} language and thus its compiler type-checks the input program to ensure no typing errors at run time. 
The typing rules for COOL, defined in the CoolAid manual, provide deduction rules for COOL expressions (e.g., if \text{x} is an integer and \texttt{y} is an integer then the expression \texttt{x + y} results in an integer).

The evaluation of a COOL program is provided using operational semantics rules.
Similar to type checking rules that deduce the types of COOL expressions, these operational semantic rules deduce values for COOL expressions (e.g.,  if \text{x} is 3 and \texttt{y} is 7 then the expression \texttt{x + y} results in 10).   These rules are also specified in the CoolAid manual.  
Fig.~\ref{fig:cool_rules} shows the type checking and operational semantics rules for applying binary operators $*,+,-,/$ to expressions $e1$ and $e2$.

\begin{figure}
\small
\begin{minipage}{0.43\linewidth}
\begin{prooftree}
\AxiomC{$\vdash e_1: Int$}
\AxiomC{$\vdash e_2: Int$}
\BinaryInfC{$op \in \lbrace *,+,-,/ \rbrace$}
\UnaryInfC{ $\vdash e_1 \; op \; e_2: Int$}
\end{prooftree}
\end{minipage}
\begin{minipage}{0.46\linewidth}
\begin{prooftree}
\AxiomC{$\vdash e_1: Int(i_1)$}
\AxiomC{$\vdash e_2: Int(i_2)$}
\BinaryInfC{$op \in \lbrace *,+,-,/ \rbrace$}
\UnaryInfC{$v_1 = Int(i_1 \; op \; i_2) \vdash e_1 \; op \; e_2:v_1$}
\end{prooftree}
\end{minipage}
\caption{Type Checking and Operational Semantic Rules}\label{fig:cool_rules}
\end{figure}

\paragraph{Reference Compiler} Students taking compiler courses using COOL often implement a full COOL compiler, which consists of main phases including lexing, parsing, type checking, semantic evaluation, and ASM transforming. To help students debug and make progress, the instructor often provides a complete ``reference'' compiler, e.g., an executable binary compiled from a COOL compiler implementation written in C or OCaml.  Students then use the reference compiler to check each step of their implementation in both output results and error messages.  \tool{} relies on the reference compiler to return error messages to the user.

\section{\tool{}}
\newcommand{\ftTextMate}{\footnote{According to Microsoft's documentation, "VS Code's tokenization engine is powered by TextMate grammars. TextMate grammars are a structured collection of regular expressions and are written as a plist (XML) or JSON files. VS Code extensions can contribute grammars through the grammar contribution point."\cite{Syntax_Highlight_Guide}}}

\newcommand{\ftLSP}{\footnote{LSP is a protocol developed by  Microsoft for  VS  Code.  This protocol defined a  standard for editors to communicate with language servers,  decouples language-specific features with the editors. It enables the reuse of Language Server among editors and reduces repeated work (implement the same feature for each editor). }}

\begin{figure}
\includegraphics[width=1\linewidth]{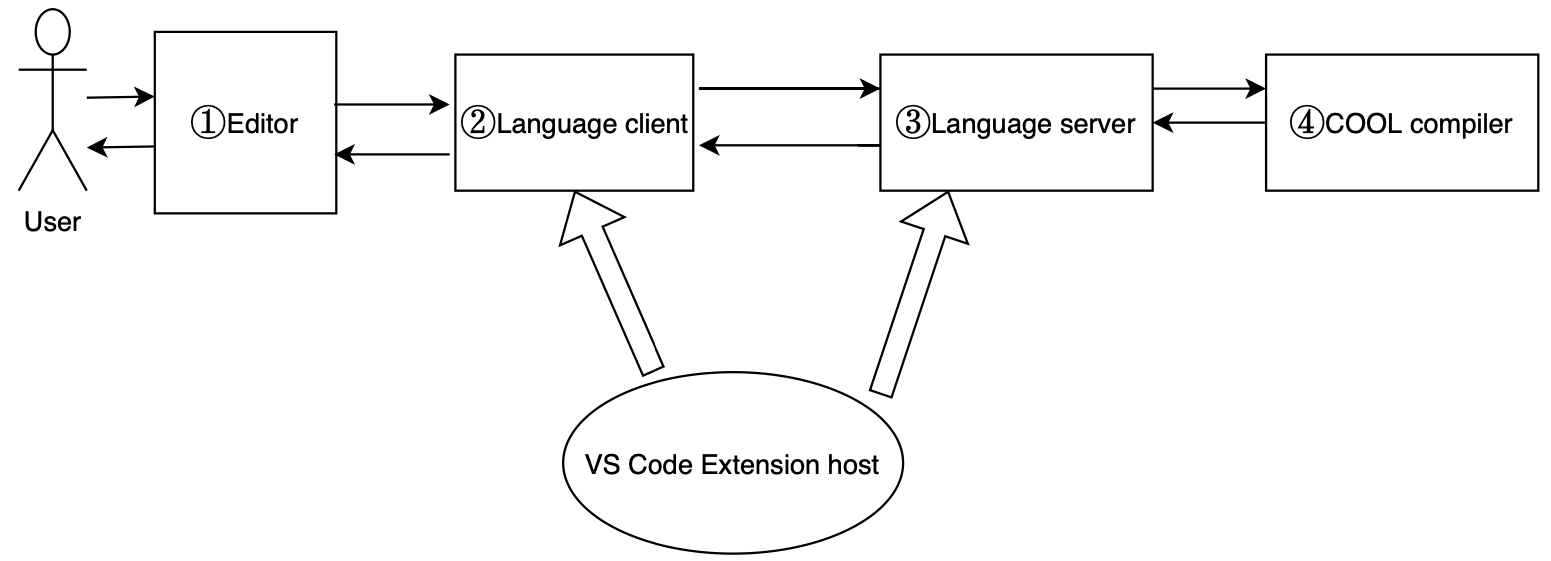}
\caption{Overview of \tool{}}
\label{fig:overview}
\end{figure}
  
The COOL Language Server Extension (\tool{}) can be found through the VSCode's extension window or its Github repository given in \S\ref{sec:intro}. It is automatically activated when the user opens a COOL file with a \texttt{.cl} extension and adds language support to the COOL program opened in the editor. 
\tool{} is composed of four components shown in Fig.~\ref{fig:overview}.

  \begin{itemize}%
      \item[\textcircled{1}] The \textbf{editor} provides a GUI interface to the user, and is the host of many editing features such as the syntax highlight and autocomplete. It also interfaces with the language client (\textcircled{2}) through its API to display information such as error messages from the COOL compiler.
  
      \item[\textcircled{2}] The \textbf{language client} interacts with the editor and the language server by sending and receiving information, e.g., getting the content of an opened file, cursor position, or display error messages. Whenever the client is activated,  it creates a language server and sends data to the server.

\item[\textcircled{3}] The \textbf{language server} is a proxy to our COOL compiler. When a request from the client is received,  the server retrieves the source code and sends it to the COOL compiler for analysis. The analysis results are composed into the LSP response format and returned to the client for display.
  
  \item[\textcircled{4}] The COOL reference \textbf{compiler} performs standard compilation phases such as parsing, type checking, and communicating results (e.g., warning and error messages) to the language server).
  \end{itemize}

\subsection{Syntax Highlight}\label{sec:syntaxhighlight}

Syntax highlight in \tool{} is handled by the VSCode editor. The editor tokenizes the source code using the regex rules specified in a \texttt{TextMate} configuration file. After that, the editor assigns colors to the tokens displayed in the IDE based on their defined roles and the colors specified in IDEs current theme (e.g., comments are blue in some themes while red in others).

\begin{figure}[h]
\centering
\begin{lstlisting}[frame=single,basicstyle=\footnotesize\ttfamily, columns=fullflexible]
   "scopeName": "source.cool",
   "name": "COOL",
   "fileTypes": ["cl"]
   "patterns":[
      "line_comment": {
         "begin": "--", "end": "$\\n?",
         "name": "comment.line.double-dash.cool"
      }
      "class": {
         "match": "[A-Z][a-z]*", 
         "name": "entity.name.type.class.cool"
      }]
\end{lstlisting}
\caption{Syntax Highlight Configuration in \tool{}}
\label{fig:texmate_syntax}
\end{figure}

For \tool{}, we create a TextMate\cite{textmate_home} JSON file consisting of regular expressions (regexes) for all of COOL's keywords, tokens, etc.  Fig.~\ref{fig:texmate_syntax} shows a snippet of a TextMate configuration having regexes for line comments and class identifiers.
Tab.~\ref{tab:regex} shows syntax highlighting regexes for several main components in COOL.

When the user opens a COOL (\texttt{.cl}) file, the \tool{} extension is activated and uses these rules for syntax highlighting.
While regexes can be complicated, the COOL reference compiler already contains similar regexes for lexing and parsing.  Thus, we adapt and reuse these regexes. %

\paragraph{Example} Fig.~\ref{fig:syntax_highlight} demonstrates syntax highlight in \tool{}. In this default \emph{VSCode Light} theme, keywords are black, control flow structures are purple, class names are cyan, integer constants are green, and so on.

\begin{table}
\caption{Syntax Highlight Rules}\label{tab:regex}
\small
\begin{tabular}{ll}\toprule
\textbf{COOL Constructs} & \textbf{Regex Descriptions}\\
\midrule
    integer   & non negative integer with no leading zero \\\hline
    ID(Class) & alphanumeric string start with a uppercase letter \\\hline
    ID(other) & alphanumeric string start with a lowercase letter \\\hline
    special ID & self, SELF\_TYPE \\\hline
    string    & alphanumeric string enclosed by double quote \\\hline
    comment   & line start with "--" and
              block enclosed by "(*", "*)" \\\hline
    keywords  &  \textit{class, else, false, fi, if, in, inherits, isvoid, let, loop,}\\
              & \textit{pool, then, while, case, esac, new, of, not, true} \\\bottomrule
\end{tabular}
\end{table}

\subsection{Auto Completion}\label{sec:autocompletion}

\tool{} supports standard autocompletion (also called "Intellisense" in VSCode\cite{intelliSense}). This allows the user to type a few keywords and the IDE can suggest and autocomplete common code structures.  For example, when \tool{} detects that the user is typing the keyword \texttt{if}, it will ask if the user wants to replace that with the COOL snippet \texttt{if condition then expression else expression fi}.  This autocompletion is straightforward yet useful in code development in an IDE as they help accomplish the goal of reducing typing effort and syntax errors from the developer.

To offer syntactically and structurally correct snippets, \tool{}'s autocompletion has knowledge about the syntax and grammar of COOL. Similar to syntax highlighting (\S\ref{sec:syntaxhighlight}), this is achieved at the editor level using a user-supplied TextMate configuration file. 
This TextMate file consists of rules mapping prefix strings (e.g., \texttt{let}) with code snippets (e.g., \texttt{let var ... in ..}). 
The code snippets can have multiple placeholders for any sub-expressions, and the user can switch between them using the “Tab” key.
Once the configuration file is loaded with the extension, the editor will automatically try to match the prefix with the string the user entered. If the input matched (can be fully or partially) the prefix of any snippets, the name and description of the snippet will be listed in a drop-down manual for the user to select.

\begin{figure}[h]
\begin{lstlisting}[frame=single,basicstyle=\footnotesize\ttfamily,columns=fullflexible]
"COOL_class_inherits": {
   "prefix": "class",
   "body": [
      "class ${1:Name} inherits ${2:Object}{",
      "\t${0:body}",
      "};"],
   "description": "COOL: class inherits"}
\end{lstlisting}
\caption{TextMate Autocompletion Configuration}\label{fig:autocompletionconfig}
\end{figure}

Fig.~\ref{fig:autocompletionconfig} shows an example of \tool{}'s TextMate configuration. Here, "body" defines the code snippet line by line, and the \texttt{"\$\{...\}"} marks the placeholders. This snippet defines the snippet for a \texttt{class} definition in COOL (i.e., when the user types \texttt{class}, the \tool{} can fill in the skeleton for defining a class in COOL). 

\begin{figure}
    \begin{minipage}{.44\linewidth}
        \includegraphics[width=1.1\linewidth]{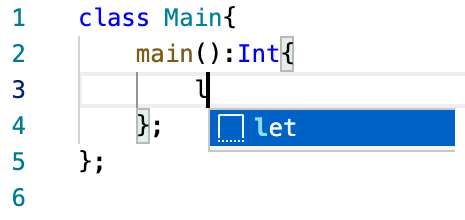}
    \end{minipage}
    \begin{minipage}{.46\linewidth}
        \includegraphics[width=1.2\linewidth]{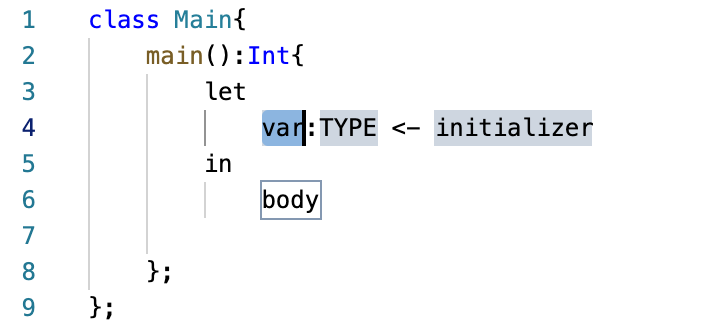}
    \end{minipage}%
    \caption{Autocompletion in \tool{}}\label{fig:autocomplete}
\end{figure}

\paragraph{Example} Fig.~\ref{fig:autocomplete} shows an example when \tool{} senses that the user wants to declare a new variable through the \texttt{let} keyword and generates the snippet for new variable declaration with appropriate placeholders for specifying the type and initial value.

\subsection{Communication with the User}\label{sec:interaction}

\tool{} uses LSP for feedback communication, i.e., displaying information from the backend COOL compiler such as errors and warnings to the frontend editor.  The design of the LSP allows the analysis to be done in separate processes, and communicate the results with the editor through inter-process communication (thus analysis done in the server does not affect the user's interaction with the editor).

\tool{} includes a pair of language client and server.  The language sever is used as a proxy to communicate with the COOL reference compiler. When the user saves their code changes, the server invokes the COOL compiler on the code and sends outputs to the client. 
The language client acts as a middle layer between the editor and the language server.
It receives data such as error messages from the server and invokes the editor's API to display them to the user.

For \tool{}, reporting data such as error messages from the COOL compiler to the client is achieved by format and translating the error message into a representation that the editor can understand and display.
More specifically, when the compiler reports an error or warning is found in the program, the editor displays the line number, the stage of compilation, and the detailed error message. These errors and warnings can come from any compilation phase, e.g., failing to parse certain expressions or having incorrect type associations.

Note that the COOL compiler used line numbers to record the error location, whereas the LSP requires an error location represented in character position. This issue was handled by the languages server, in which we replaced the line number with the position of corresponding new line characters.

\begin{figure}[h]
\includegraphics[width=0.9\linewidth]{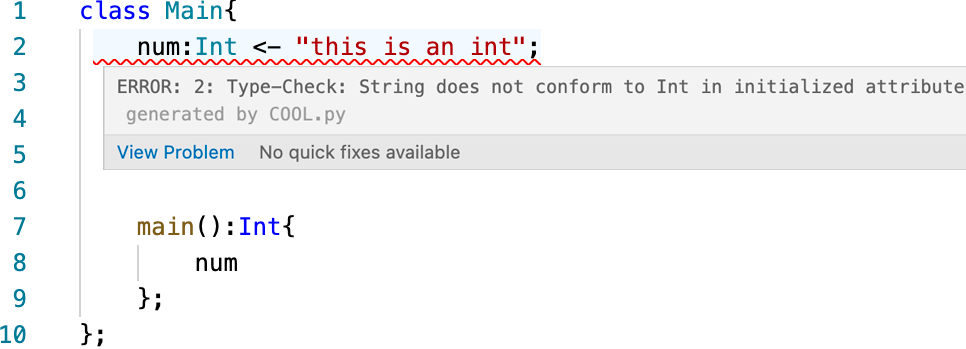}
\caption{Display an Error Message}
\label{subfig:err_msg}
\end{figure}

\paragraph{Example} Fig.~\ref{subfig:err_msg} shows how \tool{} communicates error messages to the user via the editor. In line 2, the user assigns a string constant to variable "num" of type Int, which is not allowed (since the types do not conform and COOL does not support type casting). This error is caught by the backend compiler and reported to the user. The red wavy line indicates the location of the error, and if the user hoover over the line, a detailed error message will appear. 

\section{Conclusion}
We presented \tool{}, a Language Server Protocol (LSP) that extends existing IDEs and editors to support the COOL programming language.  Similar to LSPs for other languages such as C++ and Python, \tool{} aims to make writing code in COOL easier through features such as complete syntax highlighting, intellisense code autocompletion, and rich error and communication mechanism between the backend compiler and frontend editor.

An indirect goal of \tool{} is to show researchers an easy way to make their research more visible and usable for users who are familiar with IDEs, but not with research tools, which are often written as command-line tools. Researchers can make their research tools (e.g., static or dynamic analysis tools) part of the LSP and interact with the user the same way \tool{} interacts with the COOL compiler. 
For example, in addition to creating a command line fault localization or program repair tool, the developers can also create an LSP connecting the tool with the IDE, allowing the users to interact with the tool to repair code directly through any code editor supporting LSP.
Given the popularity of IDEs such as VSCode and their powerful extension ecosystem, we believe this will make research tools more attractive to users, who are often excited about trying new extensions that are easy to install and use as in the case of in VSCode. 
\bibliographystyle{ACM-Reference-Format}
\bibliography{refs}

\end{document}